\documentstyle[prl,floats,aps,twocolumn,multicol,epsf,psfrag]{revtex}
\tighten
\begin{document}
\twocolumn[\hsize\textwidth\columnwidth\hsize\csname 
@twocolumnfalse\endcsname
\title {Analytic study of the null singularity inside spherical 
charged black holes} 
\author{Lior M. Burko and Amos Ori}
\address{
Department of Physics, Technion---Israel Institute of Technology,
32000 Haifa, Israel.}
\date{\today}

\maketitle

\begin{abstract}

We study analytically the features of the Cauchy horizon (CH) singularity
inside a spherically-symmetric charged black hole, nonlinearly perturbed by 
a self-gravitating massless scalar field. 
We derive exact expressions for the divergence rate of the 
blue-shift factors, namely the 
derivatives in the outgoing direction of the scalar field $\Phi$ and the 
area coordinate $r$. Both derivatives are found to grow 
along the contracting CH exactly 
like $1/r$. Our results are valid everywhere along the CH singularity, up to 
the point of full focusing. 
These exact analytic expressions are verified numerically.
\newline
\newline
PACS number(s): 04.70.Bw, 04.20.Dw
\end{abstract}

\vspace{3ex}
]

\section{Introduction}

In the last few years, the investigation of spinning and charged black holes 
led to a new picture of the spacetime singularity 
inside such black holes. 
According to this new picture, the Cauchy horizon (CH) evolves into 
a curvature singularity, which 
has the following two remarkable features: (i) It is 
{\it null} (rather than spacelike). (ii) It is {\it weak} (in Tipler's 
terminology \cite{Tipler}); namely, the tidal 
deformation experienced by extended physical objects is 
finite at the null singularity. 
In the case of a spinning black hole, the evidence for the 
occurrence of the null weak 
singularity has emerged from a systematic linear and nonlinear 
perturbation analysis \cite{ori92}. 
For the toy model of a spherical charged black hole, 
the main features of the singularity 
at the inner horizon were first 
deduced analytically from simplified models based 
on null fluids \cite{hiscock,poisson90,ori91}, and later 
confirmed numerically for a model with a self-gravitating scalar field 
\cite{brady95a,burko97}. 
(See also the approximate leading-order analysis in \cite{bonanno95}). 
In addition, the {\it local} existence and genericity of the 
null weak singularity 
were shown mathematically in Ref. \cite{ori96}, and more recently 
(in the framework of plane-symmetric spacetimes) 
in Ref. \cite{planar}. The compatibility of 
a null weak singularity with the constraint equations was shown in Ref.  
\cite{brady95b}.

We shall consider here the spherically-symmetric model of a 
charged black hole nonlinearly perturbed by a self-gravitating, 
minimally-coupled, massless scalar field. 
Despite its relative simplicity (compared to the analogous model of
a spinning black hole), 
no systematic analytic study of this model has been carried out so far. 
The goal of this 
paper is to present a simple analytic calculation, which may be the first step 
towards such a thorough analytic study: We quantitatively analyze 
the evolution of the divergent blue-shift factors along the contracting 
CH. It is well 
known (from both theoretical considerations and numerical simulations) 
that the singularity at the CH is characterized by finite values 
of the scalar field $\Phi$ and the area coordinate $r$. 
(The latter is also known to decrease monotonically 
with increasing affine parameter along the CH, due to 
the outflux of energy-momentum carried by the scalar field.) 
However, the 
{\it gradients} of $\Phi$ and $r$ diverge at the CH. More 
specifically, let $V$ be a ``Kruskal-like" ingoing null 
coordinate (i.e. an ingoing null coordinate for which the 
double-null metric function $g_{_{UV}}$ is finite and nonvanishing at the 
Cauchy horizon -- see below). Then, $r_{_{,V}}$ and $\Phi_{_{,V}}$ 
diverge at the CH. In this paper we shall calculate the evolution of 
$r_{_{,V}}$ and $\Phi_{_{,V}}$ along the contracting CH. We shall show, 
analytically, that the divergence rate of both entities is exactly 
proportional to  $1/r$. Our method of calculation is non-perturbative, and is 
therefore valid also in the region of strong focusing; 
however, we shall use the perturbative results (applicable at the early 
section of the CH) to determine the two overall coefficients characterizing 
the blue-shift divergence.

The paper is organized as follows. 
In section \ref{sec1} we describe the physical model of the 
self-gravitating massless scalar field on a spherical charged black hole, 
and present the field equations. 
In Section \ref{sec2} we carry out a leading-order perturbation analysis 
of $r$ (and use pervious perturbative results for $\Phi$)  
and calculate the $v$-derivatives of $\Phi$ and $r$ at the 
very early 
part of the CH (where the focusing effect is still negligible). 
Then, in Section \ref{sec3} we perform a fully nonlinear 
(and non-perturbative) calculation of these $v$-derivatives, which is valid 
everywhere along the contracting CH, up to the point of full 
focusing, where $r=0$ and the singularity becomes spacelike 
\cite{brady95a}. This nonlinear 
analysis leaves two coefficients undetermined -- one for each field -- 
and we determine these two coefficients by matching the nonlinear results 
to the linear results applicable at the asymptotically-early part of the 
CH. Our results are in excellent agreement with the numerically-obtained 
results \cite{burko97}.

\section{Physical Model and Field Equations} \label{sec1}

We consider here the model of a spherically-symmetric charged black hole, 
nonlinearly perturbed by a self-gravitating, spherically symmetric, 
minimally-coupled, massless scalar field $\Phi$ (the same model as that 
analyzed numerically in \cite{brady95a,burko97}). 
This model allows us to obtain nontrivial radiative dynamics 
while retaining the simplicity of the spherical symmetry.  

We write the general spherically-symmetric line
element in double-null coordinates,
\begin{eqnarray}
\,ds ^{2}=-f(u,v)\,du\,dv+r^{2}(u,v)\,d\Omega^{2},
\label{metric}
\end{eqnarray}
where $\,d\Omega^{2}$ is the line element on the unit two-sphere.
The energy-momentum tensor of a massless scalar field is
\begin{eqnarray}
T_{\mu\nu}^{\rm s}=\frac{1}{4\pi}\left(\Phi_{,\mu}\Phi_{,\nu}-\frac{1}{2}
g_{\mu\nu}g^{\alpha\beta}\Phi_{,\alpha}\Phi_{,\beta}\right).
\end{eqnarray}
The energy-momentum associated with the general 
spherically-symmetric free electromagnetic field is
\begin{eqnarray}
T_{\mu\nu}^{\rm em}=\frac{Q^{2}}{8\pi r^{4}}
\left(\begin{array}{cccc}
0 & f/2 &0 & 0 \\
f/2 &0 &0 &0 \\
0 &0 &r^{2} &0 \\
0 &0 &0 &r^{2}\sin^{2}\theta 
\end{array} \right) , 
\end{eqnarray}
where $Q$ is the electric charge. 

For a spherically-symmetric scalar field, 
the Klein-Gordon equation reduces to 
\begin{eqnarray}
\Phi_{,uv}+\frac{1}{r}\left(r_{,u}\Phi_{,v}+r_{,v}\Phi_{,u}\right)=0.
\label{KGEQ}
\end{eqnarray}
The Einstein field equations, 
$G_{\mu\nu}=8\pi (T_{\mu\nu}^{\rm s}+T_{\mu\nu}^{\rm em})$, include 
two evolution equations,
\begin{eqnarray}
r_{,uv}=-\frac{r_{,u}r_{,v}}{r}-\frac{f}{4r}\left(1-\frac{Q^{2}}
{r^{2}}\right)
\label{EEQ1}
\end{eqnarray}
\begin{eqnarray}
f_{,uv}&=&\frac{f_{,u}f_{,v}}{f}+f\left\{ \frac{1}{2r^{2}}\left[4r_{,u}r_{,v}+
f\left( 1-2\frac{Q^{2}}{r^{2}}\right)\right]\right.\nonumber \\
&-&\left. 2\Phi_{,u}\Phi_{,v} \right\},
\label{EEQ2}
\end{eqnarray}
and two constraint equations:
\begin{eqnarray}
r_{,uu}-(\ln f)_{,u}r_{,u}+r(\Phi_{,u})^{2}=0
\label{con1}
\end{eqnarray}
\begin{eqnarray}
r_{,vv}-(\ln f)_{,v}r_{,v}+r(\Phi_{,v})^{2}=0.
\label{con2}
\end{eqnarray}  

The form of the above line element and field equations is invariant to 
a coordinate transformation of the form $v\to\bar v(v),u\to\bar u(u)$. 
In what follows $u$ and $v$ will denote {\it generic}, unspecified, 
double-null coordinates. Below we shall 
often use specific types of null coordinates for specific calculations, 
and in order to avoid confusion we shall assign a special symbol to each of 
these specific coordinates. Thus, we shall denote the standard 
Eddington-Finkelstein null coordinates of Reissner-Nordstr\"{o}m (RN) by 
$u_e$ and $v_e$. We shall also use $U$ and $V$ to denote {\it Kruskal-like} 
coordinates, i.e. double-null coordinates which regularize the line 
element (\ref{metric}) at the inner horizon. 
In addition, in section \ref{sec3} we shall 
define two other types of ingoing null coordinates, $V_r$ and $V_\Phi$.

\section{Linear regime} \label{sec2}
Previous analytic and numerical studies have indicated that 
the geometry at (and near) 
the early part of the CH may well be described by the  background 
metric functions of the static (or stationary) black-hole solution 
plus a small metric perturbation. This is found to be the 
situation both in vacuum spinning black holes (analytically) 
\cite{ori92} and in the present 
model of a spherical charged black hole (numerically) \cite{burko97}. 
Moreover, the 
perturbations become arbitrarily small as one approaches the asymptotic past 
of the CH. In the very early part of the CH, the perturbations are dominated 
by their linear part, and the singularity is well described by the 
linear metric perturbation. In the later part of the CH, 
however, nonlinear effects become 
exceedingly important, as demonstrated, e.g., by the contraction of the CH. 

Accordingly, we shall schematically divide the CH into two parts:
\begin{enumerate}
\item The {\it linear regime}, i.e. the 
asymptotically-early part of the CH, where 
the metric perturbations (and the scalar field) are still very small, and 
a leading-order analysis is adequate; 
\item The {\it nonlinear regime}, i.e. 
the later part of the CH where the focusing (and possibly other nonlinear 
effects) become important. At the future end of the nonlinear regime 
the area of the CH shrinks to zero, and the singularity becomes spacelike. 
\end{enumerate}
In this Section we shall consider the linear regime, and obtain 
expressions for the blue-shift factors, namely, 
the $v$-derivatives of $\Phi$ and $r$. 
The nonlinear regime will be the subject of Section \ref{sec3}. 

In the linear regime we may treat $\Phi$ as a linear Klein-Gordon field 
over a fixed RN background. The evolution of such a field 
was analyzed in \cite{gursel79} and more recently in 
\cite{ori97a,ori97} (using a different method). 
For a spherically-symmetric scalar field satisfying an inverse power-law 
\begin{eqnarray}
\Phi\cong {v_{e}}^{-n}\quad\quad\quad({\rm EH})
\label{EH}
\end{eqnarray}
at the event horizon (EH), the asymptotic behavior at
the early part of the CH was found to be \cite{gursel79,ori97a} 
\begin{eqnarray}
\Phi\cong A{v_{e}}^{-n}+B{u_{e}}^{-n}\quad\quad\quad({\rm CH}),
\end{eqnarray}
where $A$ and $B$ are constants which only depend on the ratio 
$Q/M$, $M$ being the black-hole mass. 
Since we are primarily interested here in the $v$-derivative of $\Phi$, 
we only need the value of $A$, which was found in Refs. 
\cite{gursel79,ori97} to be
\begin{equation}
A=\frac{1}{2}\;\frac{r_+}{r_-}\;
\left(\frac{r_+}{r_-}+\frac{r_-}{r_+}\right) ,
\label{a}
\end{equation}
$r_{\pm}$ being the value of $r$ at the outer and inner horizons of RN.

One finds that both at the EH and at the CH 
\begin{equation}
\Phi_{,v_e}\propto{v_e}^{-p},
\label{lin11}
\end{equation} 
where $p\equiv n+1$, and
\begin{equation}
\Phi_{,v}^{\rm CH}/\Phi_{,v}^{\rm EH}\to A.
\label{lin4}
\end{equation}
Here and below the arrow denotes the limit of large advanced 
time (corresponding to $v_e\to\infty$).
Note that the last relation is explicitly gauge invariant, so it 
holds for any type of ingoing null coordinate 
$v$, and not only for $v=v_e$.

Next, we consider the $v$-derivative of $r$ at the CH. In the pure RN 
geometry, $r_{,v_e}$ dies off exponentially (in $v_e$) at the CH. 
In the presence of the self-gravitating scalar field, however,  
$r_{,v_e}$ decays as a {\it power-law} of $v_e$ (see below). In the 
asymptotically-early portion of the CH 
(the ``linear regime") which concerns us here, 
the effect of the scalar field is dominated by the second-order term 
(i.e., the term quadratic in 
derivatives of the scalar field), and higher-order 
corrections are negligible. We shall now calculate this
leading-order term of $r_{,v_e}$. 

Viewing Eq. (\ref{con2}) as a linear first-order differential 
equation for $r_{,v}$, we formally integrate it and obtain 
\begin{eqnarray}
r_{,v}(v)=-f(v)\int^{v} 
\frac{r(v')}{f(v')}\;\left[\Phi_{,v}(v')\right]^{2}\,dv'.
\label{lin2}
\end{eqnarray}
Here [and also in Eq. (\ref{lin2a}) below] the integration is done along 
lines of constant $u$, and we omit the dependence on $u$ for 
brevity. From this exact expression we 
now extract the term quadratic in (derivatives of) $\Phi$. 
Since ${\Phi_{,v}}^2$ appears explicitly in the integrand, we simply need 
to replace $f$ and $r$ in the right-hand side by the corresponding 
unperturbed metric functions of RN, which we denote
$f_{_{\rm RN}}$ and $r_{_{\rm RN}}$:
\begin{eqnarray}
r_{,v}(v)=-f_{_{\rm RN}}(v)\int^{v} 
\frac{r_{_{\rm RN}}(v')}{f_{_{\rm RN}}(v')}\;\left[\Phi_{,v}(v')\right]^{2}
\,dv' \;\;.
\label{lin2a}
\end{eqnarray}
Note that this equation is invariant to the choice of gauge for 
the coordinate 
$v$. We shall now evaluate the integral at the right-hand side, using the 
null coordinate $v_e$. 
In this gauge, $f_{_{\rm RN}}$ decays exponentially 
at the CH:
\begin{eqnarray}
f_{_{\rm RN}}\propto e^{-\kappa_- v_e},
\label{fed}
\end{eqnarray}
where $\kappa_{-}$ is the surface gravity of the inner horizon. 
On the other hand, $\Phi_{,v_e}$ decays as an inverse power of $v_e$, and 
$r_{_{\rm RN}}$ approaches a nonzero constant, $r_-$, at the CH. Therefore, 
since the relative change of $\Phi_{,v}$ (and $r_{_{\rm RN}}$) is 
exponentially slower than that of $f_{_{\rm RN}}$, 
to the leading order in $1/v_e$ we can take $\Phi_{,v}$ outside the integral, 
and substitute $r_{_{\rm RN}}\cong r_-$ [as well as Eq. (\ref{fed}) 
for $f_{_{\rm RN}}$]. Doing so, we obtain (to the leading order in 
$1/v_e$) \footnote {
In the transition from
Eq. (\ref{lin2}) to Eq. (\ref{lin2a}) we got rid of all terms of order higher
than quadratic in $\Phi$. Thus, in principle Eq. (\ref{lin2a}) should
include both the zero-order and the
second-order parts of $r_{,v}$. The zero-order term is
represented by the (implicit) integration constant in Eq. (\ref{lin2a}).
This zero-order term is exponentially small, however, and is thus
negligible compared to the quadratic term in Eq. (\ref{lin3a}).
}
\begin{eqnarray}
r_{,v_{e}}\cong -\frac{r_{-}}{\kappa_{-}}\;\left(\Phi_{,v_e}^{\rm 
CH}\right)^{2}\; .
\label{lin3a}
\end{eqnarray}
Finally, using Eq. (\ref{lin4}), we find
\begin{eqnarray}
{r_{,v_{e}}\over \left(\Phi_{,v_e}^{\rm EH}\right)^{2}}
\to -{r_{-}\over\kappa_{-}}\,A^2 .
\label{lin3a1}
\end{eqnarray}
In particular, we have in the linear regime
\begin{eqnarray}
r_{,v_{e}}\propto v_{e}^{-2p}.
\label{lin12}
\end{eqnarray}

One clarification should be made here concerning the precise meaning of the 
parameters $r_-$ and $\kappa_-$, and the coordinate $v_e$, in the perturbed 
spacetime. (Originally these entities are only defined in the pure 
RN geometry.) We know that outside the black hole, both the scalar field 
and the metric perturbations decay at late time, and the geometry approaches 
that of RN. In particular, the mass function approaches a limiting 
value $M$. We thus define $r_-$ and $\kappa_-$ according to the value of 
these parameters in the asymptotic RN geometry, i.e. according to their 
standard definition in terms of $M$ and $Q$ 
(with $M$ being the above late-time limit of the mass function; 
Note that the charge $Q$ is a fixed parameter in our model). In a similar 
way, we also define the coordinate $v_e$ with respect to this late-time 
asymptotic RN geometry. More specifically, we may define 
$v_e$ according to the affine parameter $\lambda$ along 
a line of constant $r>2M$ (or along the EH), by taking 
$v_e(\lambda)$ to be the same function as in the pure RN geometry 
(with a mass parameter $M$ defined as above). 
Note that once the entities $M$, $r_-$, $\kappa_-$, and $v_e$ 
were defined in the linear regime, their extension to the 
nonlinear regime is trivial. 

One might be puzzled by the relevance of the asymptotic external 
mass parameter to the internal dynamics near the perturbed 
CH (and particularly 
to the definition of the inner-horizon parameters $r_-$ and $\kappa_-$), 
especially when the divergence of the mass function at the 
CH is recalled. The resolution of this puzzle relays on the basic 
features of the geometry inside perturbed charged (or spinning) black holes: 
On the one hand, the geometry is drastically different from that of RN 
(or Kerr), as expressed by the divergence of curvature at the CH. On the  
other hand, the geometry is very similar to RN (or Kerr) in 
terms of the metric functions: The metric perturbations are arbitrarily small 
at the asymptotic past of the CH. [Roughly speaking, 
the divergence of curvature indicates 
the divergence of {\it derivatives} of the metric functions 
(with respect to the regular background coordinate $V$) at the CH.] 
This smallness of metric perturbations is the necessary basis for the 
entire perturbative approach: As it turns out, the 
perturbation analysis (when properly formulated) respects 
the smallness of the metric perturbations, and not the divergence of 
curvature. That is, the typical ratio of two successive terms 
in the nonlinear perturbation expansion is comparable to the 
small metric perturbations, and not to the diverging curvature (this 
is fortunate, because otherwise the perturbative approach 
would render useless). 
This was demonstrated analytically for spinning black holes \cite{ori92}, 
and numerically for charged ones \cite{burko97}. 

In the above analysis of the linear regime 
(based on the perturbative approach), $r_-$ and $\kappa_-$ appear as  
{\it parameters of the background RN geometry}, and their definition should 
therefore be based on the asymptotic mass function $M$. 
On the other hand, the divergence of the mass function 
(whose definition also involves the {\it derivatives} of $r$) 
at the perturbed CH merely 
reflects the divergence of $r_{_{,V}}$ there, due to the 
{\it perturbation} (which undergoes infinite blue-shift). 
Obviously, this divergence has no relevance to the background parameters 
$r_-$ and $\kappa_-$.

\section{Nonlinear regime}\label{sec3}
We turn now to analyze the divergence rates of $r_{,v}$ and $\Phi_{,v}$ 
along the nonlinear, strong-focusing, portion of the CH. Here, it will 
be insufficient to calculate the leading-order perturbations, so we 
must carry out a full nonlinear calculation.

We shall base our calculation on two assumptions:
\begin{enumerate}
\item \label{1} 
For an appropriate choice of coordinates $u,v$, the 
line-element (\ref{metric}) is valid up to the singular CH, and both 
functions $f$ and $r$ are finite and nonvanishing along the singular CH. 
We shall denote such regular 
coordinates by $U,V$, and refer to them as {\it Kruskal-like} 
coordinates. (Of course, the choice of $U$ and $V$ is not unique.) 
We shall also set $V=0$ at the CH.
\item \label{2}
There exists at least a single outgoing null geodesic, $u=u_0$, which 
intersects the CH and which satisfies the following two requirements:
\begin{enumerate}
\item \label{2a}
Along $u=u_0$, $r$ and $\Phi$ are monotonic functions of $v$ 
in a neighborhood of the CH, 
\item \label{2b}
Along $u=u_0$, both $r_{_{,V}}$ and 
$d\Phi/dr$ (i.e. $\Phi_{_{,V}}/r_{_{,V}}$) diverge at the CH.
\end{enumerate}
\end{enumerate}
The validity of assumption \ref{1} is strongly supported by the perturbative 
approach, at least in the early part of the CH. 
Moreover, recent numerical simulations \cite{burko97} confirm its validity 
in the entire CH up to the point of full focusing (where the singularity 
becomes spacelike).
\footnote{
If assumption \ref{1} were valid only in a portion of 
the CH, then the analysis below would nevertheless 
be applicable to this portion (provided 
that the outgoing null ray considered in assumption \ref{2} intersects 
this portion).
} 
Assumption \ref{2} is justified, because at least in the asymptotically-early 
part of the CH, Eqs. (\ref{lin11},\ref{lin12}) ensure the 
required monotonic behavior, and also imply 
$d\Phi/dr\propto {v_e}^p\to\infty$. 
In addition, in the linear regime 
the standard ingoing Kruskal-like coordinate, $V\equiv e^{-\kappa_- v_e}$, 
regularizes the line element at the CH, and satisfies 
$r_{_{,V}}\propto {v_e}^{-2p}\,e^{\kappa_- v_e}\to\infty$. 
We can thus take $u_0$ to be in this asymptotically-early section 
(in fact, the numerical 
simulations \cite{burko97} confirm that the asymptotic relations 
of assumption 2 hold everywhere along the CH).

To analyze the evolution of $r_{,v}$, we shall use the evolution equation 
(\ref{EEQ1}), viewing it as a first-order ordinary equation for $r_{,v}$. 
Our goal is to integrate this equation in the ingoing direction, 
along the CH. This integration 
would be trivial if the last term on the 
right-hand side (which couples this equation to the other evolution 
equation) were absent. Fortunately, on approaching the CH, this last term 
becomes arbitrarily small compared to the preceding one. 
For example, in a Kruskal-like $V$, the first term in the right-hand 
side diverges (at least at 
$u=u_0$), whereas the second one is finite. (Note that although each 
of these terms depends on the gauge, their ratio 
is gauge-invariant.) This suggests that, when 
integrating this equation along the CH, the last term could be dropped. 
In order to analyze this equation in a more systematic and 
elegant way, we define a new ingoing 
null coordinate $V_r$ in the neighborhood of the CH, 
by 
\begin{eqnarray}
V_r (v)\equiv r(u=u_0,v),
\label{vr}
\end{eqnarray}
and reexpress Eq. (\ref{EEQ1}) in terms of $V_r$. 
To transform $f\equiv -2g_{uv}$ from $v$ to $V_r$, 
we first calculate $g_{_{UV_r}}$:
\begin{eqnarray}
g_{_{UV_r}}=g_{_{UV}}/ \left( {dV_r/ dV} \right)=
g_{_{UV}}/ \left( {dr/ dV} \right)_{u_0}.
\label{guv}
\end{eqnarray}
Since $g_{_{UV}}$ is finite (assumption 1), and 
$\left( {dr/ dV} \right)_{u_0}$ diverges (assumption 2b), we find that 
$g_{_{UV_r}}$ vanishes everywhere along the CH, and so is $g_{_{uV_r}}$. 
Defining $z(u)\equiv \left(r_{_{,V_r}}\right)_{_{\rm CH}}$, 
Eq. (\ref{EEQ1}) now reduces to the trivial equation 
$z_{,u}=-(r_{,u}/ r)\; z$.
Its general solution is 
\begin{eqnarray}
z=C/r,
\label{z}
\end{eqnarray}
where $C$ is an integration constant. 
Note that this exact equality holds {\it everywhere} along the CH.
Calibrating $C$ at $u=u_0$, we find 
\begin{eqnarray}
r_{_{,V_r}}={r_0\over r}\left(r_{_{,V_r}}\right)_{u_0}={r_0\over r}
\quad\quad\quad 
{\rm (CH)}, 
\label{rvr}
\end{eqnarray}
where $r_0$ is the $r$-value of the CH at $u=u_0$. 
The first of these two equalities has an explicit gauge-invariant form, 
so we can immediately transform it to a generic gauge and write it as 
\begin{eqnarray}
{r_{,v}\over \left(r_{,v}\right)_{u_0}}\to{r_0\over r}
\label{nl1}
\end{eqnarray}
(later we shall use this result for $v=v_e$).

The analysis of the evolution of $\Phi_{,v}$ proceeds in a similar way. 
This time we use the KG equation (\ref{KGEQ}), 
viewed as an ordinary differential equation 
for $\Phi_{,v}$, and integrate it along the CH. By virtue of assumption 
\ref{2b}, the second term in the parentheses in Eq. (\ref{KGEQ}) 
is negligible at the CH 
(at least at $u=u_0$) compared to the preceding one. To make an optimal 
use of this fact, we transform Eq. (\ref{KGEQ}) from $v$ to 
the new ingoing null coordinate 
\begin{eqnarray}
V_\Phi (v)\equiv \Phi(u=u_0,v),
\label{vphi}
\end{eqnarray}
defined in a neighborhood of the CH. The last term in the transformed 
equation is proportional to $r_{_{,V_\Phi}}$. But 
\begin{eqnarray}
r_{_{,V_\Phi}}=r_{_{,V_r}}\left(dV_r /dV_\Phi\right)=
r_{_{,V_r}}\left(dr/d\Phi\right)_{u_0}.
\label{rvphi}
\end{eqnarray}
At the CH, $r_{_{,V_r}}=C/r$ and $\left(dr/d\Phi\right)_{u_0}\to 0$ 
(assumption \ref{2b}), so the last term in the transformed 
equation (\ref{KGEQ}) vanishes. 
Defining $y(u)\equiv \left(\Phi_{_{,V_\Phi}}\right)_{_{\rm CH}}$, 
Eq. (\ref{KGEQ}) becomes 
$y_{,u}=-(r_{,u}/ r)\; y$,
whose general solution is 
\begin{eqnarray}
y=K/r,
\label{y}
\end{eqnarray}
where $K$ is an integration constant. 
Calibrating $K$ at $u=u_0$, we find 
\begin{eqnarray}
\Phi_{_{,V_\Phi}}={r_0\over r}\left(\Phi_{_{,V_\Phi}}\right)_{u_0}
={r_0\over r}\quad\quad\quad {\rm (CH)},
\label{phivphi}
\end{eqnarray}
and again, the first equality may be immediately transformed to a generic 
gauge: 
\begin{eqnarray}
{\Phi_{,v}\over\left(\Phi_{,v}\right)_{u_0}}\to{r_0\over r}.
\label{nl2}
\end{eqnarray}

We shall now match the non-linear results (\ref{nl1}) and (\ref{nl2}) 
to the leading-order results at the linear regime. 
To that end, we take our reference outgoing ray $u=u_0$ to be in the 
asymptotically-early section of the CH. We can then use 
the results of the previous section [e.g. Eqs. (\ref{lin4}) 
and (\ref{lin3a1})] for $\left(r_{,v}\right)_{u_0}$ and 
$\left(\Phi_{,v}\right)_{u_0}$, and also substitute $r_0=r_-$. Combining Eq. 
(\ref{nl1}) (with $v=v_e$) and Eq. (\ref{nl2}) with 
Eqs. (\ref{lin3a1}) and (\ref{lin4}), respectively, we obtain
\begin{equation}
{\Phi_{,v}\over\Phi_{,v}^{\rm EH}}\to{r_-\over r}A
\label{nl3}
\end{equation}
and
\begin{eqnarray}
{r_{,v_{e}}\over\left(\Phi_{,v_e}^{\rm EH}\right)^{2}}
\to-{{r_-}^2\over r\kappa_{-}}\,A^2.
\label{nl4}
\end{eqnarray}
These exact relations hold everywhere along the CH. 

More explicitly, for initial data $\Phi\cong {v_{e}}^{-n}$ at 
the EH, the asymptotic behavior at the CH is (to leading order in $1/v_e$)
\begin{eqnarray}
\Phi_{,v_e}\cong -n{r_-\over r}A\,{v_e}^{-(n+1)}\;\;,\;\;
r_{,v_e}\cong-n^2{{r_-}^2\over r\kappa_{-}}\,A^2\,{v_e}^{-2(n+1)}.
\label{n21}
\end{eqnarray}
These results take an especially simple form when expressed in terms 
of $\Psi\equiv r\,\Phi$ and $r^2$:
\begin{equation}
\Psi_{,v_e}\cong-nr_- A\,{v_e}^{-(n+1)}\;\;,\;\;
\left(r^2\right)_{,v_e}\cong-2n^2{{r_-}^2\over \kappa_{-}}
\,A^2\,{v_e}^{-2(n+1)}.
\label{n22}
\end{equation}
(Note that to the leading order in $1/v_e$, which concerns us here, 
the contribution of $r_{,v_e}$ to $\Psi_{,v_e}$ is negligible.) 
That is, to the leading order in $1/v_e$, the $v$-derivatives of $\Psi$ 
and $r^2$ at the CH are {\it independent of $r$} (and $u$). 
The translation of the above results from $v_e$ to any other type of 
ingoing null coordinate (e.g. $V$) is straightforward.

The above  results are verified numerically in Ref. \cite{burko97}. 
The terms at the two sides of Eqs. 
(\ref {nl3}) and (\ref {nl4}) are evaluated numerically along 
an outgoing null ray that intersects the strong-focusing portion of the CH.
The numerical results are in excellent agreement with the above analytic 
prediction.

It would be an interesting challenge to try 
generalizing these results to the CH singularity of a generic spinning 
vacuum black hole.

This research was supported in part by the United States--Israel 
Binational Science Foundation, and by the Fund for the Promotion of 
Research at the Technion.



\end{document}